\documentclass{article}
\usepackage{subcaption}
\usepackage[english,activeacute]{babel}
\usepackage{amsmath,amsfonts,amssymb,amstext,amsthm,amscd}
\usepackage{latexsym}
\usepackage{graphicx}
\usepackage{subfig}
\usepackage{float}
\usepackage{dcolumn}
\usepackage{bm}

\usepackage{here} 
\usepackage{fancyhdr}
\usepackage{physics}
\usepackage{multicol}
\usepackage[utf8]{inputenc}
\usepackage{pgf-spectra}

\usepackage{fancyhdr}
\usepackage[paperwidth=8.5in, paperheight=11.0in, top=1.0in, bottom=1.0in, left=1.0in, right=1.0in]{geometry}

\begin{document}
\fancypagestyle{plain}{
   	\renewcommand{\headrulewidth}{1pt}
   	\renewcommand{\footrulewidth}{1pt}
}

\renewcommand{\footrulewidth}{1pt}
\renewcommand{\tablename}{Tabla}


\title{Quantum Cryptography Using Momentum and Position Variables in a Simple Optical Arrangement.}
\author{\small{Alejandro Arias Jimenez \footnote{a24027978, alejandro.ARIAS-JIMENEZ@etu.univ-amu.fr} , Baptiste Sevilla \footnote{s18000139, baptiste.sevilla@etu.univ-amu.fr}}\\		
	   \small{Aix Marseille Universite}}
\date{\small{April 15, 2025}}

\maketitle

\begin{abstract}

In this work, we explore an experimental implementation of quantum key distribution (QKD) using position and momentum quantum states. By employing a setup that includes a laser, a slit, and lenses to generate a Fourier transform, we demonstrate a variation of the BB84 protocol. Our technique shows how a system with these characteristics can be implemented within a quantum framework.

\end{abstract}

\begin{multicols}{2}

\section{Introduction}

After the first results of quantum mechanics were recognized by the physics community, researchers began exploring practical applications of these phenomena, one of which is quantum cryptography. In 1984, Charles Bennett and Gilles Brassard proposed an algorithm for securely distributing a cryptographic key over a quantum channel, enabling secure communication through a classical channel. (Bennett, C. H., $\&$ Brassard, G, 2014)

The BB84 algorithm is based on the effects of measurement in quantum mechanics. When we measure a state in quantum mechanics, the wave function describing the state will collapse into only one of the eigenfunctions of the operator (Sakurai $\&$ Napolitano, 2017).

If Alice sends a prepared state on a certain basis, Bob would be able to read it if he measures it on the corresponding basis. If Bob uses the wrong basis, then his result is not determined by Alice's preparation and is therefore not useful. When Bob measures the qubits, he will decide on a basis, and when he is done, Alice will communicate the basis she used to send the qubits, and they will discard the ones that were measured incorrectly. In this scenario, if a third person, Eve, interacts with the system, she would be detected because the system could collapse depending on her measurement. Due to the no-cloning theorem, she can't avoid this risk. 

The no-cloning theorem states that there is no unitary operator that satisfies the following condition. 

$$U \ket{\psi} \ket{0}=\ket{\psi} \ket{\psi} $$

Since a qubit cannot be cloned without perturbing it, Eve would never be able to copy Alice’s qubits, send them to Bob, and wait for the basis to be published before measuring the qubits. If Eve interacts with the system, some of the qubits that Bob measures in the correct basis will not match Alice’s preparation, providing clear evidence that someone is tampering with their communication.

In this work, our goal is to design an experimental setup to achieve quantum key distribution using a laser, a slit to encode classical bits, and the Fourier transform to prepare the state in two different bases. In the standard basis, we measure position, while in the Fourier basis, we measure momentum. This approach works because the position and momentum operators do not commute, as stated by the uncertainty principle:

$$[\hat{x},\hat{p}] = i \hbar$$

This work is based on the article by (Lemelle et al., D. S., Almeida et al., but with our modifications to the algorithm and the optical system.

In optics, the Fourier transform allows us to decompose a wave into its spatial frequency components. This mathematical tool helps us model the transformation that light undergoes when passing through a lens. Additionally, in quantum mechanics, the Fourier transform relates the wave function in position space to momentum space. The Fourier transform is defined as follows: 

$$ F[U(x)] = \frac{1}{2\pi} \int_{-\infty}^{\infty} U(x)e^{ikx} dx $$

In this work, we applied this principle to prepare a state in the momentum basis, allowing us to use quantum principles for key distribution. An important detail to mention is that, in this system, we use a green laser without an optical fiber, which introduces some classical behaviours. For example, in a purely quantum experiment, photons would carry either momentum or position information, but in the classical approach, light is treated as a wave that contains information in both bases simultaneously.

The objective of this experiment is to obtain an optical arrangement to demonstrate how we can safely distribute a key using the momentum and position experiment. 

\subsection{Experimental Set Up }\label{Introduccion}     

We reproduce the experimental setup from the article \cite{article2} with some modifications. 

\begin{figure}[H]
    \centering
    \includegraphics[width=0.45\textwidth]{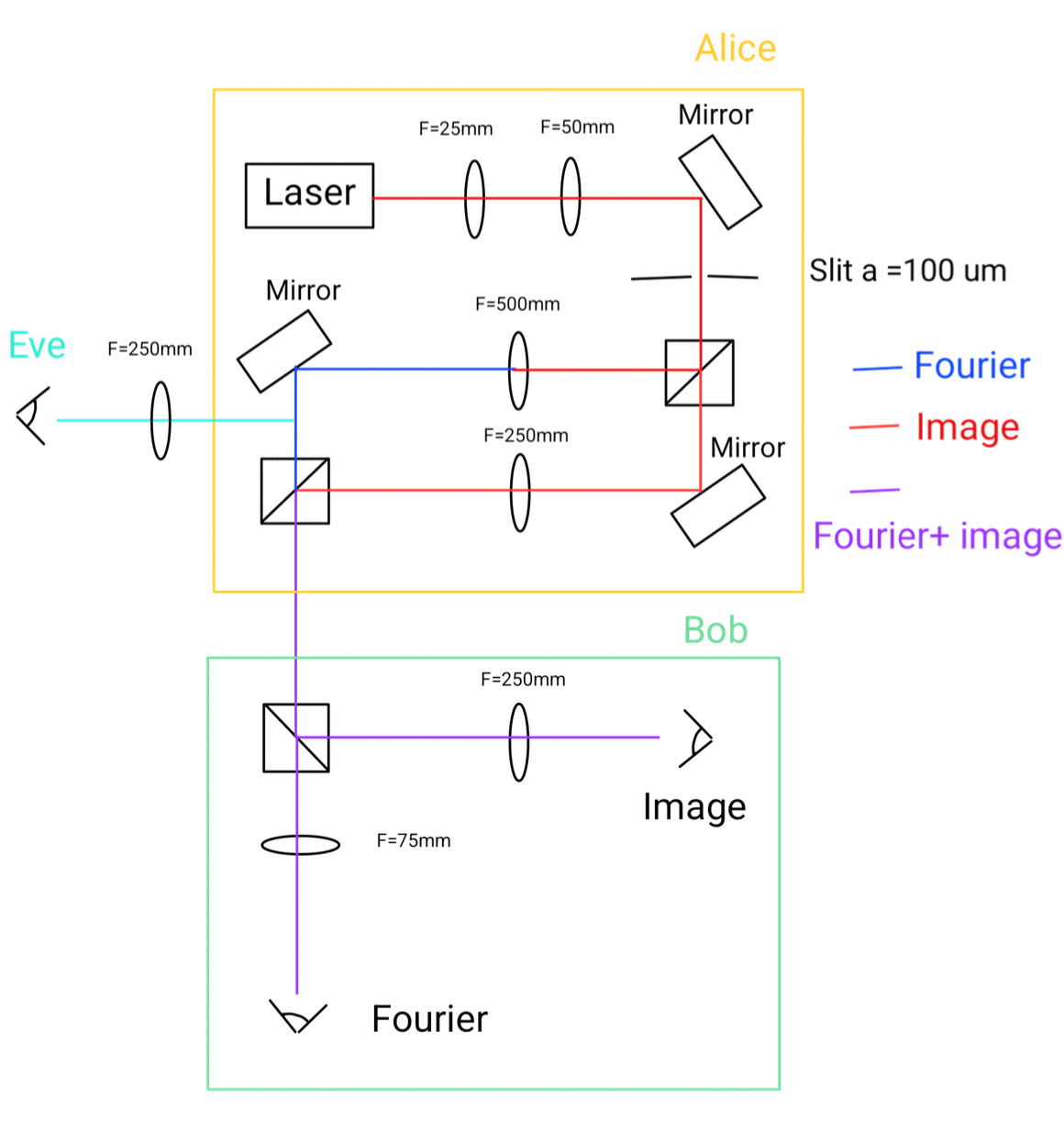}
    \caption{Scheme of out experimental set up.}
\end{figure}

We use one laser with a wavelength of $532nm$, then we use two lenses to expand the beam before it passes through the slit. Indeed, we want to have a large beam to cover the aperture of the slit. The slit aperture is $100\mu m$.Then the light passes through a beam-splitting cube which splits the light into two beams. One is the image, whereas the second is the Fourier. With this configuration, we can have two bases. However, they do not have the same optical path. The image has a lens with a focal length of $250mm$ , whereas the Fourier has a lens with a focal length of  $500mm$. When we reunite these beams using mirrors, we have a small phase shift. We will use this phenomenon to encode our bits. The second beam-splitting cube combines the two beams, then we have a third cube to split the light into two beams again. One is the image, and the other is the Fourier.

This array will allow us to obtain a variant of the BB84 QKD algorithm. In this case, if we measure in the incorrect basis, we don't have any information about the system and can't perform any measurement. This is the main difference from the classical BB84 algorithm. In this case, Eve's perturbation will be detected by measuring in the basis that Alice prepared the state and not being able to determine any information about the system.

More specifically, every 5 seconds, Alice sends a state prepared in one of the basis and containing the classical bit. If Bob measures in the correct basis, we will see a line moving from bit 1 to bit 0 or staying in that position, depending on what classical bit Alice wants to send. In the case where Bob measures in the incorrect basis, we will only see a static point and won't be able to determine anything from the system. 
When Eve tries to measure the system, she has two possibilities: measuring in the correct basis and therefore not perturbing the system, or the inverse. For example, if Eve measures in the momentum basis $p$ a prepared state in the position basis $x$ and Bob measures it in the correct basis $p$, he won't be able to determine anything about this state. If he communicates this with Alice via a classical channel, she will know that the connection isn't safe since the basis are correct.

Here we present a detailed step by step algorithm used in this experiment. 

\begin{enumerate}
    \item Alice communicates with Bob via a classical channel, sending the time interval between each bit.
    \item Alice starts to send the bits via the quantum channel on the basis she chooses and records them. 
    \item Bob decides how to measure each bit he is receiving on a random basis and records them. 
    \item Once all the bits have been distributed, Alice and Bob communicate via classical channel the basis in which they measure each bit (they do not share the bit, only the basis).
    \item Alice and Bob erase all the bits that were measured in different basis. 
    \item Alice communicates to Bob the first 10 bits that she sends and Bob compares his results. 
    \item If Bob sees that in one of the bits in which the basis coincide, he could retrieved any information then he can know that someone is tampering with the communication and he tells Alice that the connection is not secure. If he can get all of the information from all of the bits, he can tell Alice that the connection is secure and use the other bits as the key.  
\end{enumerate}

\section{Results}

We have two detectors, one for the image and the other for Fourier. In both cases, we observe two signals. One line and one dot.
\begin{figure}[H]
    \centering
    \includegraphics[width=0.35\textwidth]{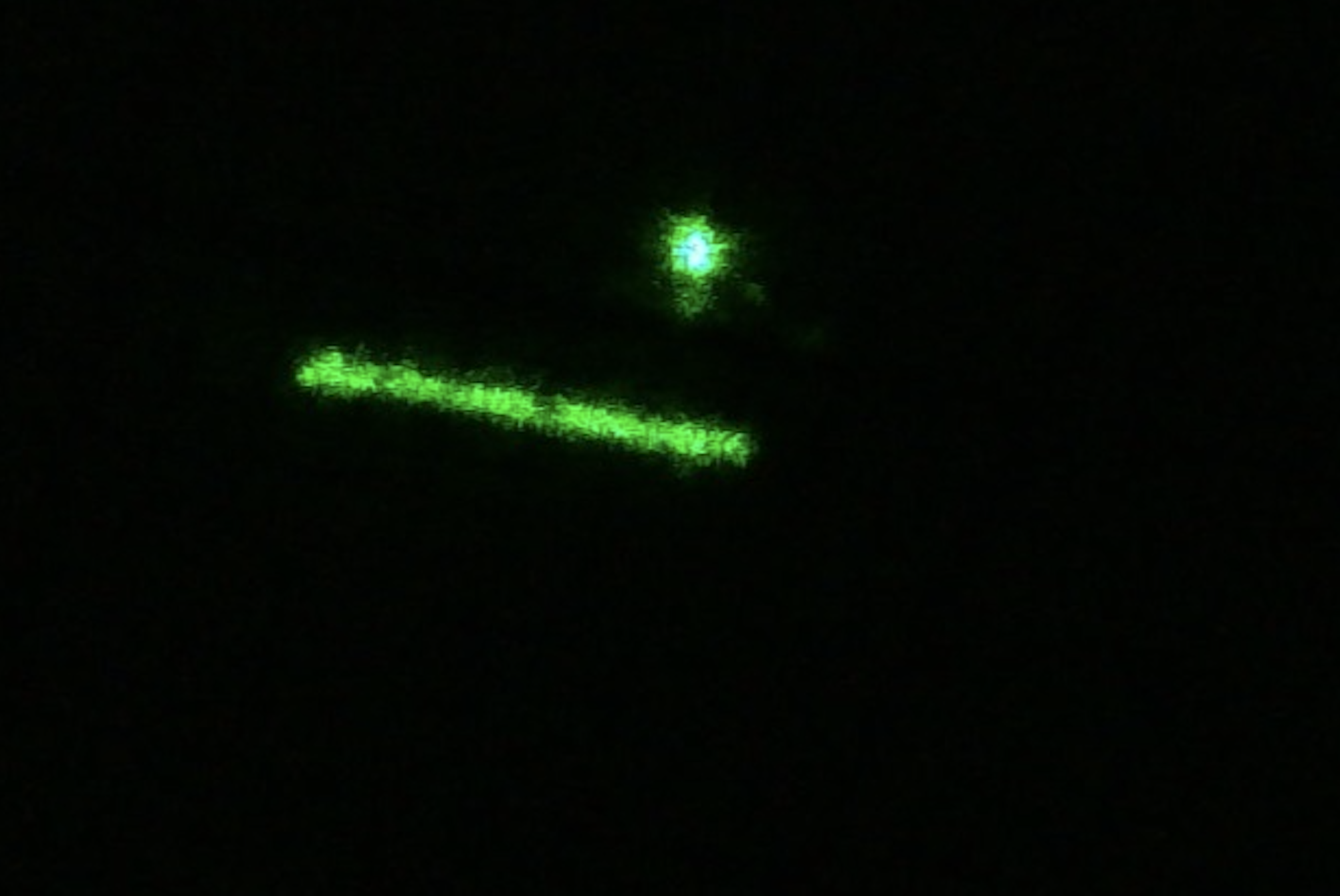}
    \caption{Dot and line}
\end{figure}

We need to block one path (Fourier or image) to have some results. If we block the Fourier path, we have the dot on the image screen and the line for the Fourier screen. When the normal path is blocked it is the inverse. 

\begin{figure}[H]
    \centering
    \includegraphics[width=0.45\textwidth]{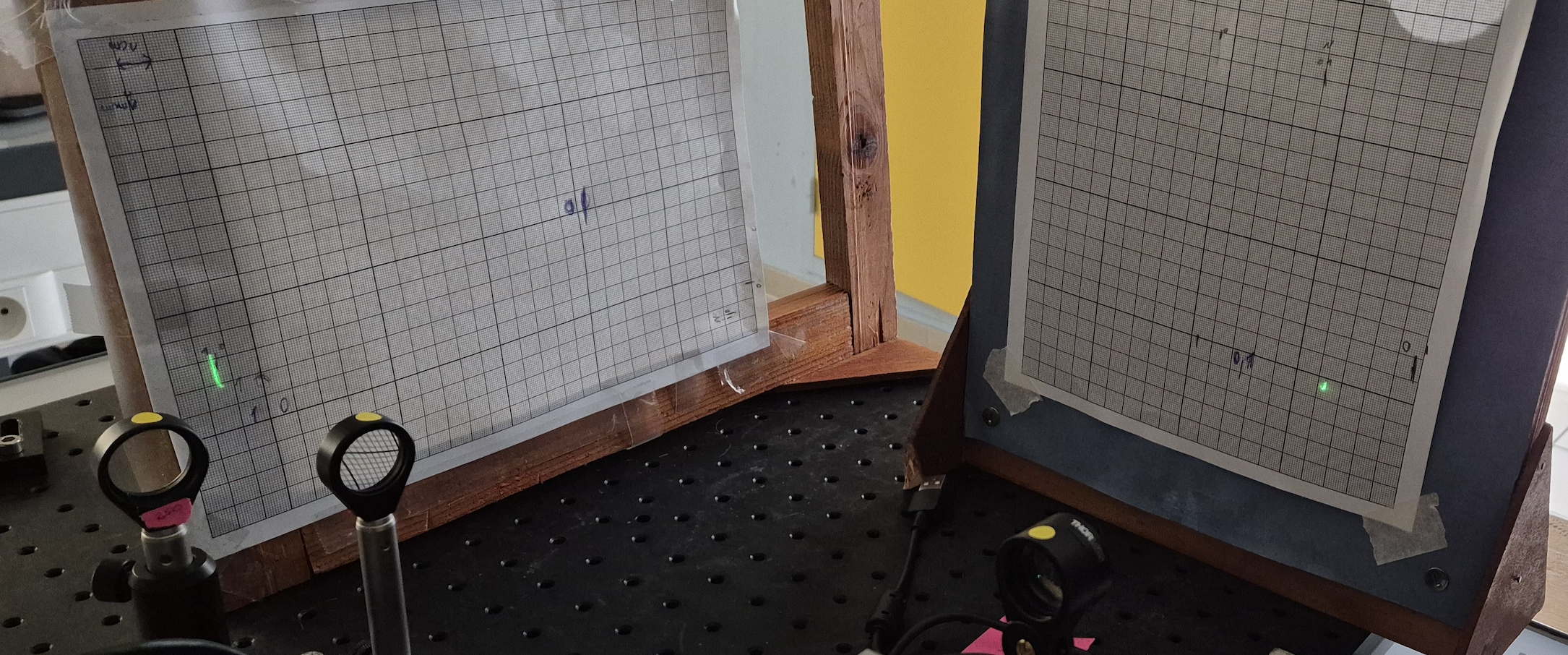}
    \caption{Dot on Fourier screen and line on Image screen, Fourier path blocked}
    \includegraphics[width=0.45\textwidth]{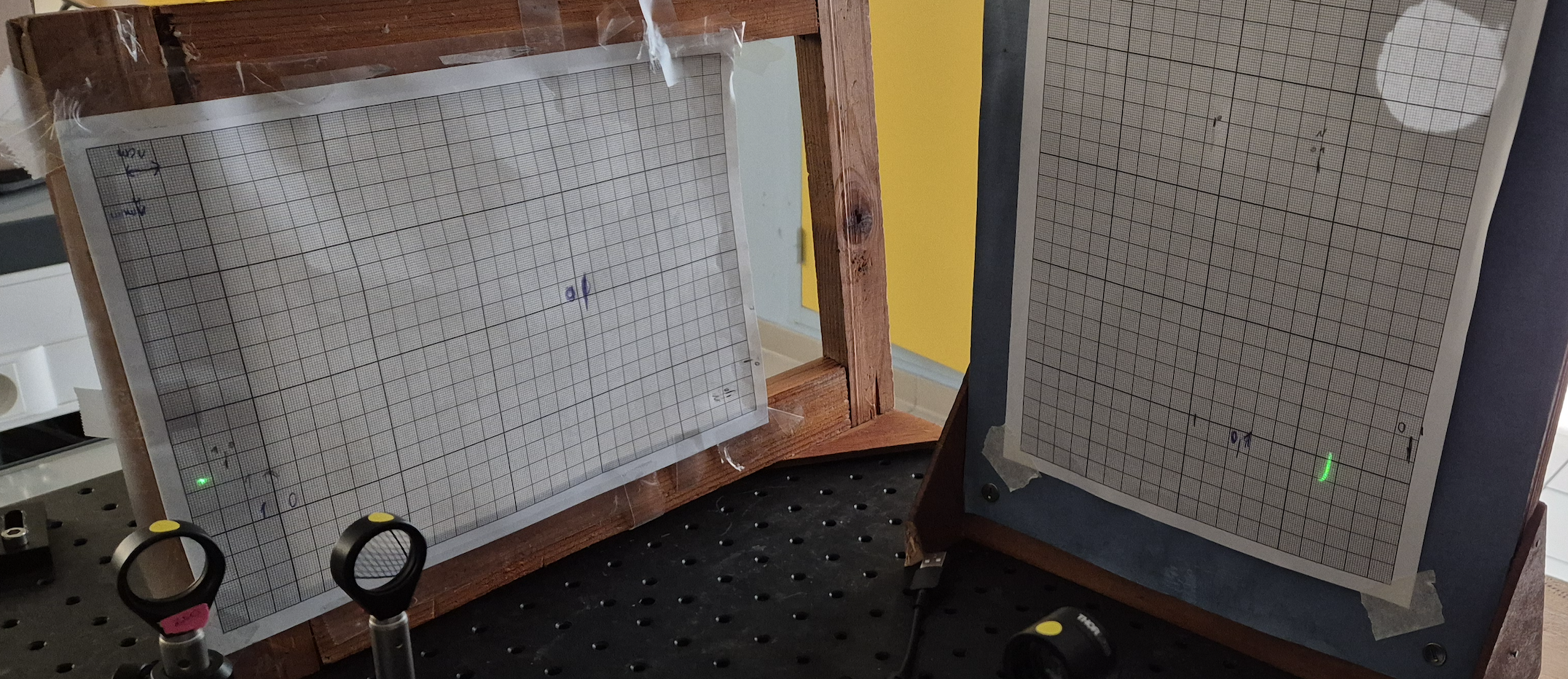}
    \caption{Dot on Image screen and line on Fourier screen, Normal path blocked}
\end{figure}

In our set up to do cryptography we need to move the slit. We can adjust the position of the slit. We only want two positions, one corresponds to the bit 0 and the other corresponds to the bit 1. We put the bit 0 with the maximum intensity where the beam passes entirely through the slit. For the bit 1 we move to $2mm$ from the position 0. We need a displacement large enough to clearly observe the shift on the screen, but we also need to ensure that the beam still passes through the slit. After few tests $2mm$ is a good distance for our set up. 

\subsection{Fourier blocked}

\subsubsection{Fourier screen}
We block the Fourier path and we observe a dot on the Fourier screen. When we have a dot it means we won't be able to see any movement between position 0 and position 1. 

\begin{minipage}{0.245\textwidth}
        \centering
        \includegraphics[width=\linewidth]{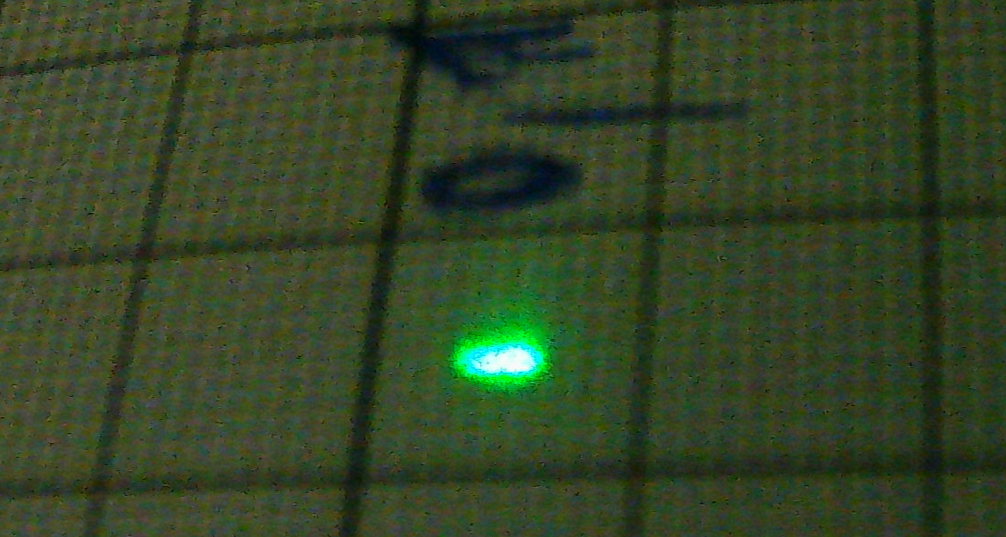}
        \captionof{figure}{Position 0}
    \end{minipage}
    \hfill
    \begin{minipage}{0.22\textwidth}
        \centering
        \includegraphics[width=\linewidth]{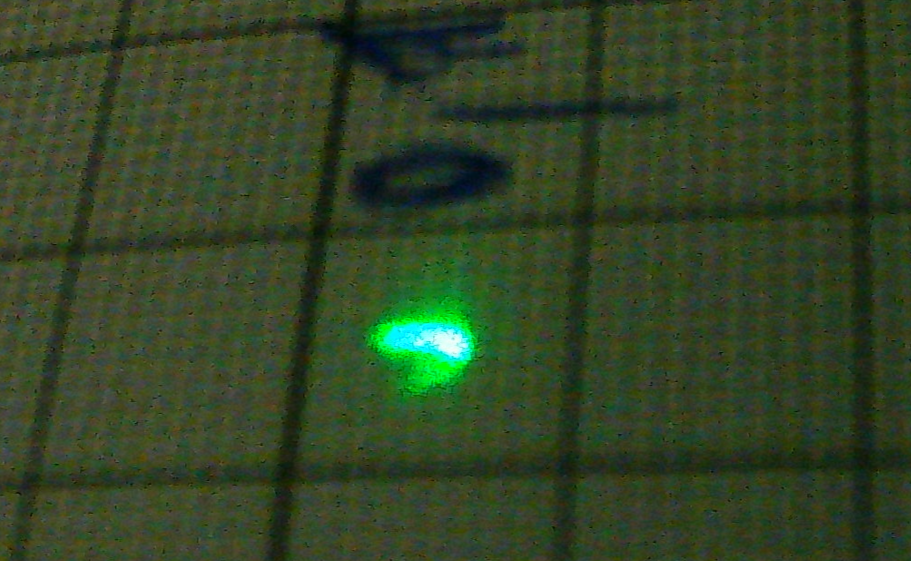}
        \captionof{figure}{Position 1}
    \end{minipage}

\subsubsection{Normal screen}
In this configuration we have a line. When we have a line we should see two different positions of this one for the position 0 and the position 1. We observe this phenomenon clearly. 

\begin{minipage}{0.24\textwidth}
        \centering
        \includegraphics[width=\linewidth]{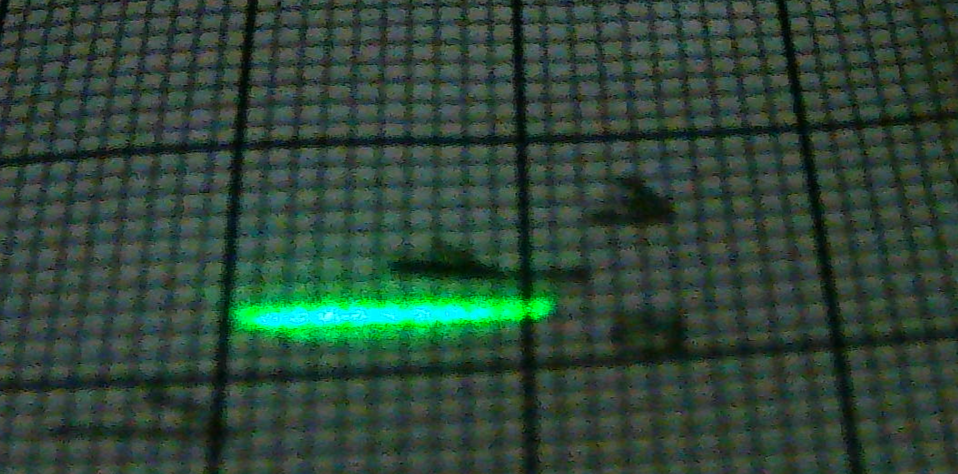}
        \captionof{figure}{Position 0}
    \end{minipage}
    \hfill
    \begin{minipage}{0.22\textwidth}
        \centering
        \includegraphics[width=\linewidth]{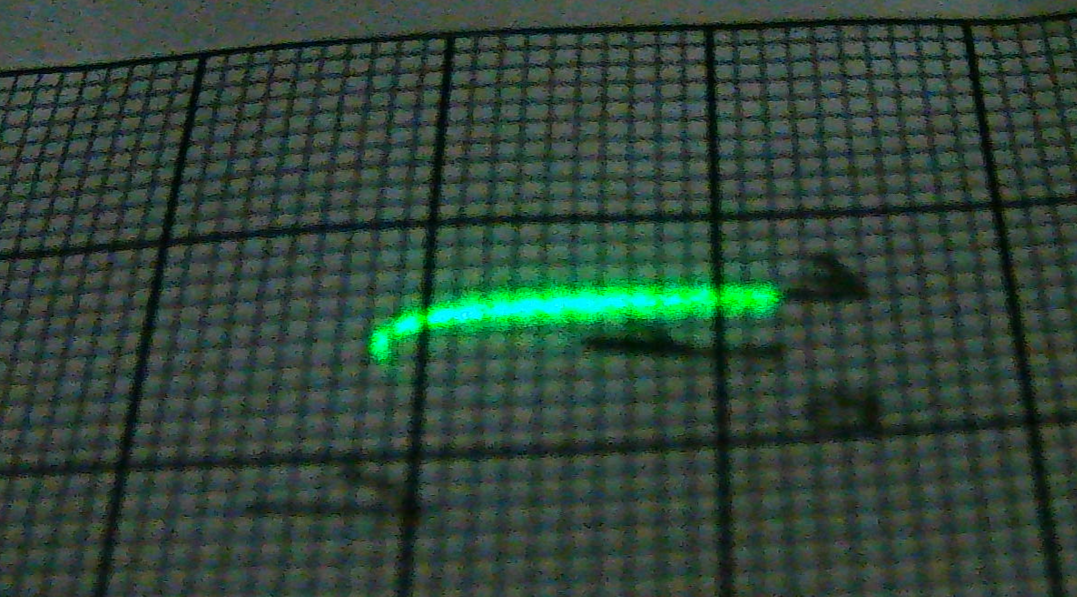}
        \captionof{figure}{Position 1}
    \end{minipage}

\subsection{Normal blocked}

\subsubsection{Fourier screen}
We have a line which moves when we switch position 0 to position 1 

\begin{minipage}{0.2\textwidth}
        \centering
        \includegraphics[width=\linewidth]{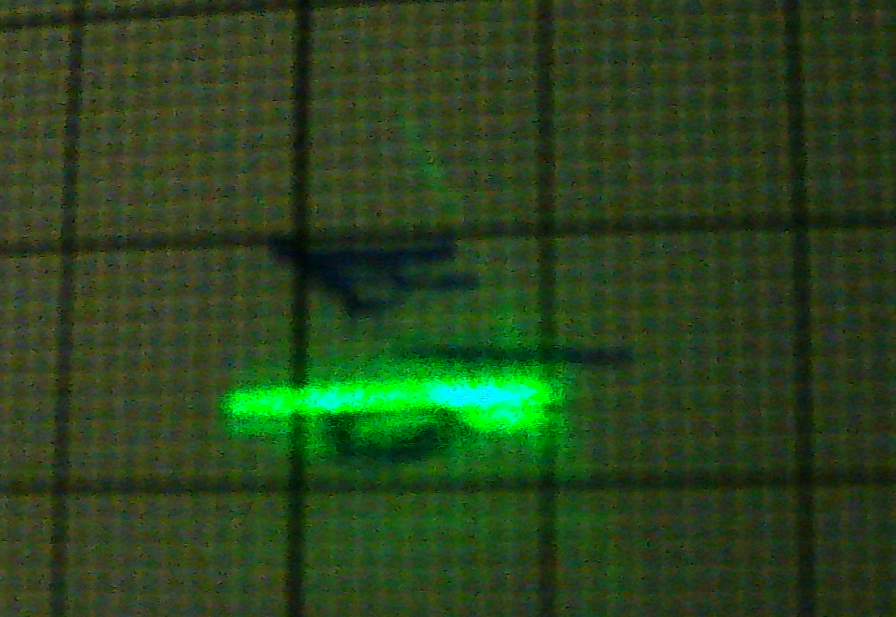}
        \captionof{figure}{Position 0}
    \end{minipage}
    \hfill
    \begin{minipage}{0.25\textwidth}
        \centering
        \includegraphics[width=\linewidth]{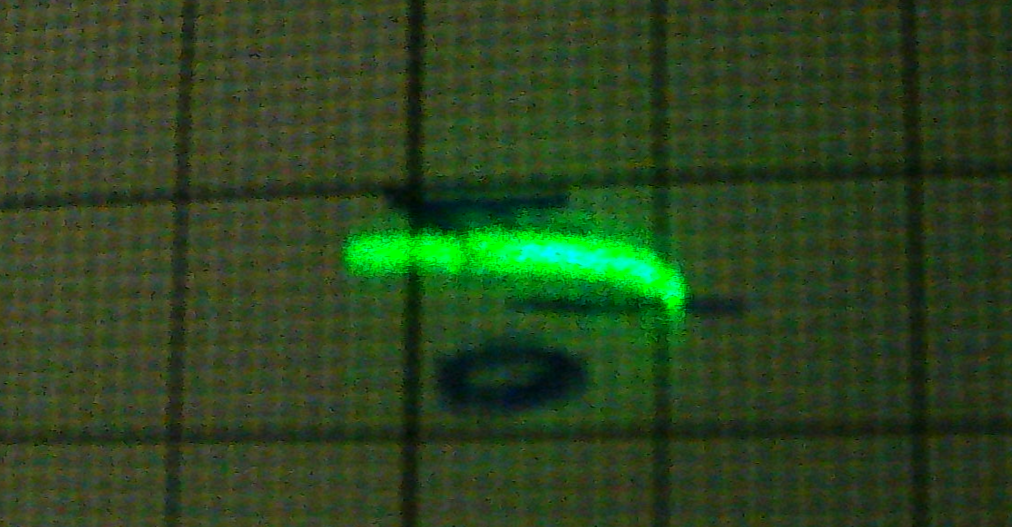}
        \captionof{figure}{Position 1}
    \end{minipage}

\subsubsection{Normal screen}
We have a dot which doesn't move when we switch positions. 

\begin{minipage}{0.25\textwidth}
        \centering
        \includegraphics[width=\linewidth]{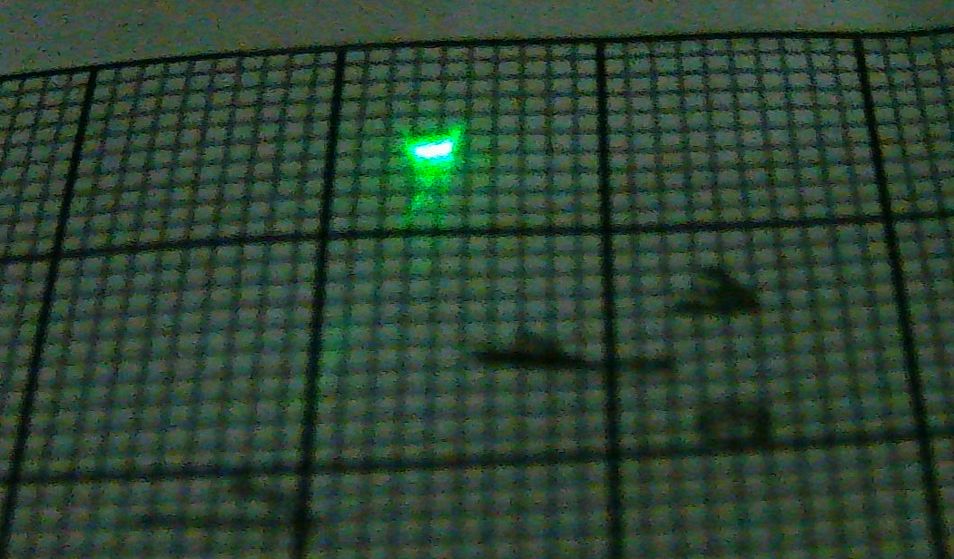}
        \captionof{figure}{Position 0}
    \end{minipage}
    \hfill
    \begin{minipage}{0.2\textwidth}
        \centering
        \includegraphics[width=\linewidth]{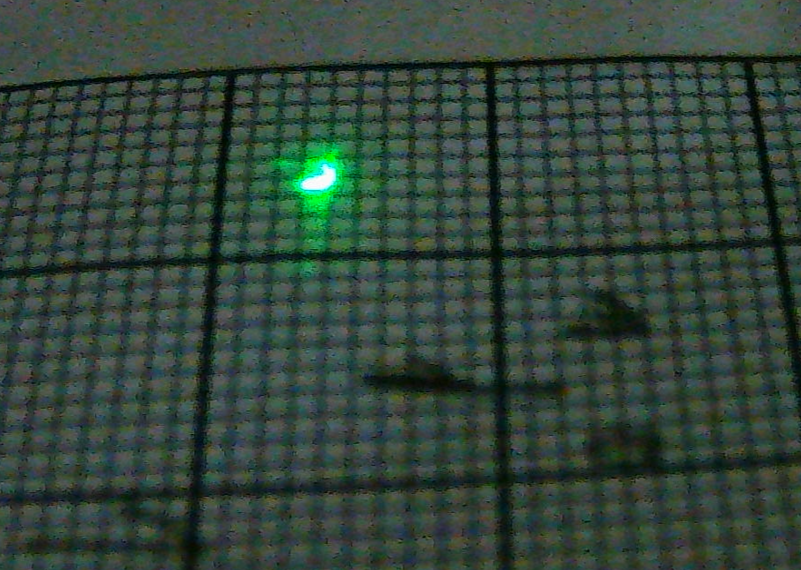}
        \captionof{figure}{Position 1}
    \end{minipage}

With this configuration we can see that when measuring in the correct basis we can obtain information about the bit but when we are measuring in the incorrect basis no information can be gathered from the system because we only see point that is not moving. It is important to mention that we can actually see a really small movement in the point since we are still in a classical framework. 

\subsection{Quantum key distribution}

The main objective of quantum cryptography is to be able to safely distribute a key in order to establish communication. 

In our experiment we used the algorithm detailed previously that differs from other algorithms and experiments. Another important detail to mention is how Eve impacts the system. 
There are 4 different scenarios with Eve, in the first one she measures on the correct basis and she is not detected. In the second and third one she measures one of the basis (either correct or incorrect) but Bob measures in the incorrect basis and they erase the bit. The last one is when she measures the incorrect basis and Bob measures the correct basis. As we can see this means that Eve has a $1/4$ probability of being noticed for only one event but in practice Alice and Bob send a lot of bits and therefore the probability of noticing Eve grows quickly. 
Therefore, Eve is represented in our system as a lens (Fourier transform) in only one of the arms, this represents when Eve is measuring in the Fourier space, with two possibilities, measuring correct or incorrect. The other arm shows the missing two possibilities for Eve.

\subsection{Write "Hello"}
We chose a basis and we look to the screen associated with this basis. The choice of the basis is abstract and we have the same results if we take one basis or the other. We chose to work with the Fourier basis it means we block the Normal path and we need to look to the Fourier screen because we saw previously in this configuration we have a change between position 0 and 1. First of all we need to convert the world "Hello" in binary then we obtain the sequence: 01101000 01100101 01101100 01101100 01101111. We observe that there is no clear distinction between bits 0 and 1, and sometimes the same bit is repeated. To address this, we introduce a time interval between two successive bits: if there is no change in position during this interval, it indicates that the current bit is the same as the previous one. We chose a waiting time of 5 seconds. After setting this convention, Alice sends the message using the slit, and Bob reads the message accordingly.

\section{Conclusions}

We successfully developed an optical setup that demonstrates how a quantum experiment can securely distribute a key using position and momentum operators. As part of our implementation, we also distributed a key to illustrate how the protocol works in practice.

Based on our findings, we identified several key differences compared to the article on which we based our work. First, our optical scheme differs notably: we replaced the screens and removed one of the lenses, as both screens were producing the same results in our setup. Additionally, the article lacks a detailed explanation of the algorithm used, and it does not clarify how two data points were obtained for each case. In contrast, in our experiment, we observed a line when measuring in the correct basis (indicating the bit value) and a dot when measuring in the wrong basis, which carried no usable information—just as expected in a quantum scenario.

To address these concerns and validate our approach, we held a meeting with Dr. Marcelo Pereira, during which we discussed the experiment he conducted nearly 20 years ago. In this discussion, Dr. Pereira explained that they used an additional lens to magnify the image and obtain a clear signal in the Fourier plane, as their setup originally produced a very small line. They also incorporated a filter to eliminate diffraction effects. Another key difference was the type of camera used in their system, the camera employed was capable of tracking light pixel by pixel and positioned very close to the image. In contrast, our setup used standard cameras placed farther from the image, which may have affected the resolution and clarity.

Despite these differences, Dr. Pereira confirmed that our experimental setup and execution were correct, and that our results are valid, as we successfully observed the quantum cryptography phenomena. He also suggested possible improvements, such as using smaller slit movements and carefully verifying the angle and alignment of the lenses, to reduce noise and achieve a clearer quantum pattern. 

Another important aspect to consider is the comparison between this technique and others, such as polarization-based experiments. A key advantage of using position and momentum operators is that the system can operate with a standard laser as the source and optical fibres to achieve genuinely quantum behaviour. Furthermore, since position and momentum are continuous variables, it is possible to transmit a larger number of bits per unit of time. This approach also tends to be more cost-effective and easier to replicate using standard telecom components.

However, there are also drawbacks. These systems are more sensitive to noise, and the protocols and algorithms required are more complex. Additionally, momentum-position-based QKD systems are generally less suitable for long-distance applications compared to discrete-variable schemes. (Weedbrook, C., Pirandola, S, 2012)

In conclusion, our results are promising. With this replicable setup, we demonstrate a practical implementation of quantum cryptography using position and momentum variables. We reached the same fundamental conclusion as the work that inspired ours: it is possible to build a quantum cryptographic system using simple components. However, our approach differs in the optical arrangement and our work goes further into the algorithmic and practical aspects of key transmission.

\end{multicols}

\end{document}